\documentclass[prx,twocolumn,floatfix,a4paper,superscriptaddress]{revtex4}

\usepackage{bm,color,amsmath,txfonts}
\usepackage{graphicx}
\usepackage{siunitx}
\usepackage{subfigure}
\usepackage{verbatim}
\usepackage{dcolumn}
\usepackage{bm}
\usepackage{epsf}
\usepackage{xcolor}
\usepackage{hyperref}
\usepackage{hhline}
\usepackage{float}
\usepackage{enumerate}
\usepackage{bbm}
\usepackage{lipsum}
\usepackage{mathrsfs}
\usepackage{amssymb}
\usepackage{ulem}

\begin{document}
	\title{Preparing magnonic non-Gaussian states by adding a single magnon onto Gaussian states}
	\author{Zi-Xu Lu}
	\author{Huai-Bing Zhu}
	\author{Xuan Zuo}
	\author{Jie Li}\thanks{jieli007@zju.edu.cn}
	\affiliation{Zhejiang Key Laboratory of Micro-Nano Quantum Chips and Quantum Control, School of Physics, and State Key Laboratory for Extreme Photonics and Instrumentation, Zhejiang University, Hangzhou 310027, China}

	\begin{abstract}
		Quantum magnonics based on YIG spheres provides a new arena for observing macroscopic quantum states. Here we propose to prepare two kinds of non-Gaussian magnonic states by adding a single magnon onto two Gaussian states, namely, coherent and thermal states. We adopt an optomagnonic system of a YIG sphere and use fast optical pulses to weakly activate the magnon-induced Stokes scattering. Conditioned on the detection of a polarized single photon, a single magnon can be added onto an initial Gaussian state. We use a microwave cavity to prepare an initial magnon coherent state and finally read out the generated single-magnon added coherent or thermal state. Both the non-Gaussian states of a magnon mode in a large-size YIG sphere are macroscopic quantum states, which exhibit many nonclassical properties, such as sub-Poissonian statistics, quadrature squeezing, and a negative Wigner function. Further, both states show a smooth transition from a quantum state to a classical state by changing the coherent or thermal magnon excitations and thus can be used for the fundamental study of the quantum-to-classical transition.   
	\end{abstract}

	\maketitle

	\section{Introduction}
	
	Non-Gaussian states (NGSs) are a type of states that can not be expressed as any convex mixture of Gaussian states~\cite{Genoni}. They manifest many novel and unique features that break the laws of classical physics~\cite{Wal} and can thus be exploited to explore many fundamental issues of quantum mechanics, such as the quantum-to-classical transition. Representative NGSs are Fock states, Schr\"odinger cat states, excitation-added coherent states~\cite{Agarwal91,Agarwal92}, excitation-subtracted squeezed states~\cite{Dakna,Wenger}, and NOON states~\cite{Afek}, among others.  These nonclassical states find a wide range of applications in quantum sensing~\cite{LF},  quantum metrology~\cite{Giovannetti11}, quantum information processing~\cite{Opatrn,Ourjoumtsev}, etc.  The experimental realization of NGSs started from microscopic systems, such as photons~\cite{Sci04,PRA05,PRA07}, atoms~\cite{Monroe,Johnson}, trapped ions~\cite{Meekhof,Roos}, hybrid atom-light systems~\cite{Hacker}, and atomic ensembles~\cite{Pezz}. Nonetheless, the preparation of NGSs in macroscopic systems, such as a massive mechanical oscillator, has been proven to be more challenging.  The radiation pressure has been exploited in optomechanics~\cite{RMP14} to prepare various NGSs of mechanical motion, e.g., by transferring NGSs from optics to mechanical motion~\cite{Khalili}. They can also be generated conditionally based on the detection of scattered photons, including single-phonon states~\cite{Simon17,TJK14}, cat-like states~\cite{Tang,Qiu20}, phonon-added coherent~\cite{Jie18R} and thermal~\cite{Enzian21,Patel21} states, etc.  Besides, recent experiments indicate that, by directly coupling to a superconducting qubit, NGSs of a massive mechanical resonator can be achieved, including Fock~\cite{Yiwen18} and cat~\cite{Yiwen23} states.

	In recent years, hybrid systems based on magnons (the quanta of spin waves) in yttrium-iron-garnet (YIG) spheres have attracted great attention~\cite{Nakamuraapp,Yuan,Bauer,Zuo24}.  The magnonic system shows two distinct advantages: its resonance frequency can be tuned in a wide range by varying the external bias field, and it can coherently couple with various quantum systems, including microwave photons~\cite{Huebl,Nakamura14,Zhang14}, optical photons~\cite{NakamuraPRB,Nakamura16,Nakamura18,Zhang16,Haigh16,Haigh21}, phonons~\cite{ZhangSA,JieL,Davis21,Shen,Dong}, and superconducting qubits~\cite{NakamuraSci15,NakamuraSA,NakamuraSci20,Xuda}. In terms of quantum magnonics, the generation of magnonic NGSs is one of the central topics. To date, a number of proposals have been offered for preparing magnonic NGSs, including Fock states~\cite{Xuda,Bittencourt}, squeezed even Fock states~\cite{Kamra}, Schr\"odinger cat states~\cite{Qiong,Sharma21,Kounalakis,CLi,PBLi}, path-entangled states~\cite{Jiang21}, NOON states~\cite{Jing}, etc.  Since the YIG sphere contains a large number of spins, the above magnonic NGSs are macroscopic quantum states.

	Here, we provide a complete scheme to prepare two kinds of magnonic NGSs, namely, magnon-added coherent (MACS) and thermal states (MATS), using an optomagnonic system, i.e., a YIG sphere, placed inside a microwave cavity.  We first prepare the magnon mode in a coherent or thermal state, and then very weakly activate the optomagnonic Stokes scattering, i.e., the magnon-induced Brillouin light scattering (BLS), which yields pairs of single photons and magnons. A single magnon is therefore added onto the initial coherent or thermal state, conditioned on the detection of a single photon. Finally, we use the microwave cavity to read out the magnonic NGS, which is mapped to the output field of the cavity. We show that the above magnonic NGSs exhibit diverse nonclassical properties, including sub-Poissonian statistics, quadrature squeezing, and a negative Wigner function.  We further analyze the impact of various parameters, such as the coherent amplitude, thermal occupation, and state reading efficiency, on those nonclassical behaviors.

	The paper is organized as follows. In Sec.~\ref{II}, we introduce the system used to prepare two kinds of magnonic NGSs in our protocol. In Sec.~\ref{III}, we explicitly show how the single-MACS (MATS) can be generated by adding a single magnon onto the magnon coherent (thermal) state, and discuss their nonclassical and non-Gaussian properties. In Sec.~\ref{IV}, we show how these magnonic NGSs can be read out by using a microwave cavity and measuring the output field of the cavity. Finally, we summarize the work in Sec.~\ref{V}.

	\begin{figure}[t]
		\includegraphics[width=0.95\linewidth]{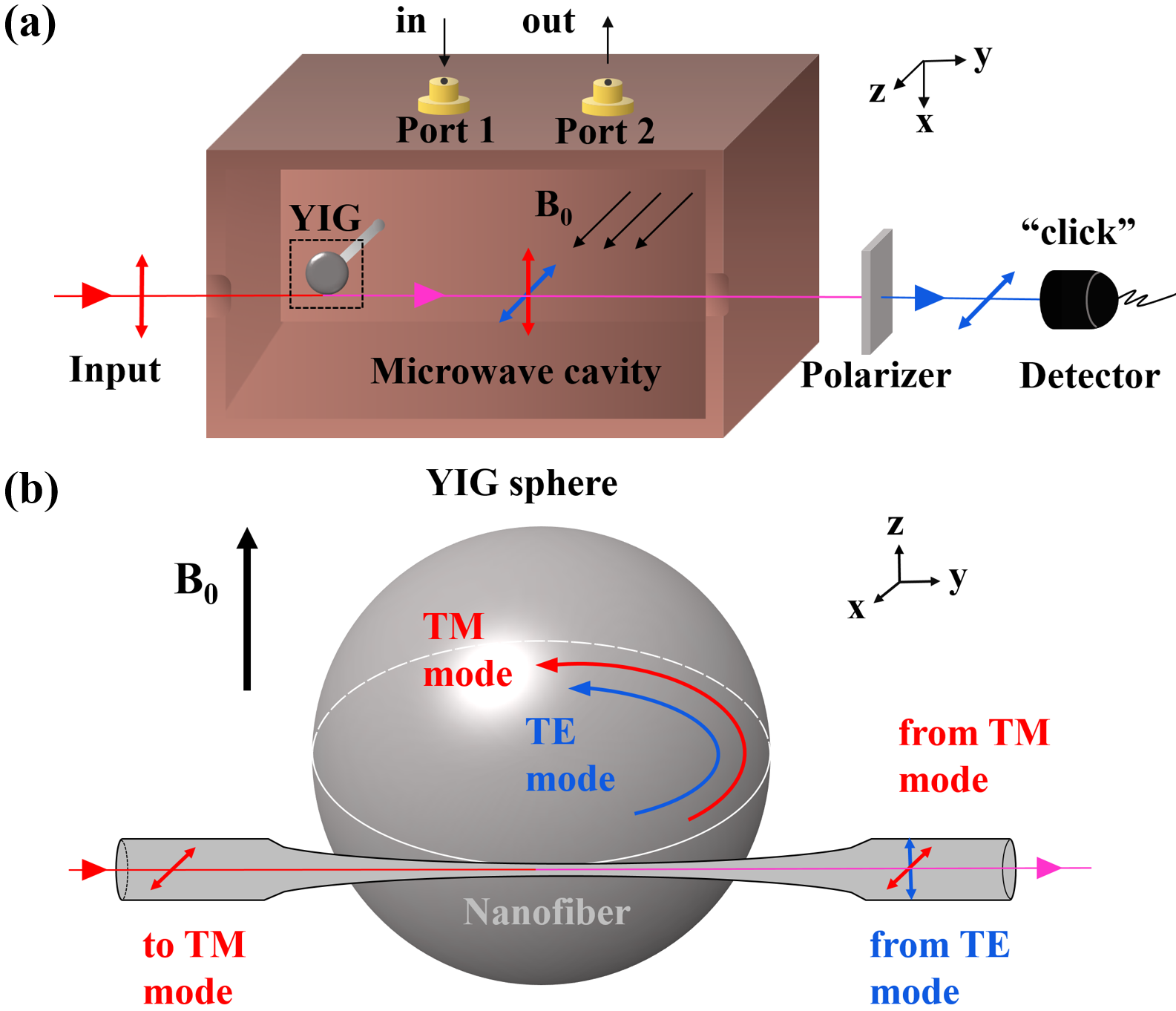}
		\caption{(a) Schematic of the system. A YIG sphere, representing an optomagnonic system, is placed inside a microwave cavity.  A WGM of the YIG sphere is driven by an optical pulse with a certain polarization via a nanofiber and the Stokes photons scattered by magnons, after passing through a polarizer, are detected by a single-photon detector. The microwave cavity is used to prepare an initial magnon coherent state and finally read out the generated magnonic NGSs.  (b) The optomagnonic system of a YIG sphere supporting a magnon mode and two WGMs with different polarizations, i.e., the TM- and TE-polarized modes. The optomagnonic interaction is manifested as the magnon-induced Brillouin light scattering.}
		\label{fig1}
	\end{figure}

	\section{The system}\label{II}

	The protocol involves an optomagnonic system, i.e., a YIG sphere, and a microwave cavity, as depicted in Fig.~\ref{fig1}(a). The YIG sphere supports both a magnetostatic mode~\cite{Gurevich}, e.g., the Kittel mode, and two optical whispering gallery modes (WGMs) with different polarizations (Fig.~\ref{fig1}(b)). Due to the magnon-induced BLS, the photons in a WGM are scattered by lower-frequency magnons (typically in gigahertz), yielding two optical sidebands of which the frequencies with respect to the WGM equal to the magnon frequency. The scattering probability is maximized when the triple-resonance condition is fulfilled~\cite{Nakamura16,Nakamura18,Zhang16,Haigh16,Haigh21}, i.e., the scattered photons enter another WGM of the YIG sphere. Due to the selection rule~\cite{Sharma17,PAP17,Nakamuranjp,Haigh18} imposed by the conservation of the angular momenta of WGM photons and magnons, the BLS exhibits a pronounced asymmetry in the Stokes and anti-Stokes scattering strength. Such asymmetry has been exploited in a number of quantum protocols~\cite{Bittencourt, Qiong,Sharma18,Jiang21,Xie22,Jie21X}. In addition, the selection rule causes different optical polarizations of the two WGMs, e.g., the transverse-magnetic (TM)- and transverse-electric (TE)-polarized WGMs in Fig.~\ref{fig1}(b).  The YIG sphere is placed inside a three-dimensional microwave cavity and the magnon mode further couples to a microwave cavity mode via the magnetic dipole interaction. The microwave cavity is used to displace the magnon mode to an initial coherent state and read out the target states that we aim to prepare.

	The Hamiltonian of the whole system can be written as
	\begin{equation}
		H=H_{0}+H_{1}+H_{2},
	\end{equation}
	where $H_{0}=\hbar\omega_{m}m^{\dagger} m$ is the free Hamiltonian of the magnon mode, $m$ ($m^{\dagger}$) is the annihilation (creation) operator, and $\omega_{m}$ is its resonance frequency which is tunable by changing the bias magnetic field $B_0$ (Fig.~\ref{fig1}(b)).  The Hamiltonian $H_{1}$ accounts for the free Hamiltonian of two WGMs, the optomagnonic interaction, and the optical drive, i.e.,~\cite{Jie21X}
	\begin{equation}
		\begin{split}	
			H_{1}/\hbar= &\omega_{1}a^{\dagger}_{1}a_{1}+\omega_{2}a^{\dagger}_{2}a_{2} +i E_{j}\big(a_{j}^{\dagger}e^{-i\omega_{p_{j}}t}- {\rm H.c.} \big) \\
			&+ g_{\rm om} \big(a_{1}^{\dagger} a_{2} m^{\dagger} +a_{1} a_{2}^{\dagger} m \big),\\
		\end{split}
	\end{equation}
	where $a_{j}$ ($a_{j}^{\dagger}$), $j=1,2$, are the annihilation (creation) operators of the WGMs and $\omega_{j}$ are their resonance frequencies, which satisfy the relation $\omega_{j}\gg\omega_{m}$ and the triple-resonance condition $|\omega_{2}-\omega_{1}|=\omega_{m}$. 
	The optomagnonic interaction is a three-wave process and the single-photon optomagnonic coupling rate $g_{\rm om}$ is typically weak~\cite{Nakamura16,Nakamura18,Zhang16,Haigh16,Haigh21}. However, the effective optomagnonic coupling strength can be significantly improved by strongly driving either the TM- or TE-polarized WGM, and $E_{j}=\sqrt{2P_{j}\kappa_{j}/\hbar\omega_{p_{j}}}$ corresponds to the coupling strength between the $j$th WGM  (with decay rate $\kappa_{j}$) and the drive field with frequency $\omega_{p_{j}}$ and power $P_{j}$. Here, without loss of generality, we assume the $a_{1}$ ($a_{2}$) mode to be the TE (TM) mode of a certain WGM orbit.  We also assume that the magnon-induced BLS occurs only between the TM and TE modes with the same WGM index, i.e., the orbital angular momentum of the WGM photons is conserved~\cite{Nakamura16,Nakamuranjp}. In this case, the frequency of the TM mode is higher than the TE mode, $\omega_{\rm TM} > \omega_{\rm TE}$, due to the geometrical birefringence~\cite{Nakamura18,Zhang16}.  

	The Hamiltonian $H_{2}$ describes the free Hamiltonian of the microwave cavity, the cavity-magnon interaction, and the microwave drive, given by~\cite{Nakamura14,Zhang14}
	\begin{equation}\label{cmBS}
		H_{2}/\hbar= \omega_{c}c^{\dagger}c+ g_{\rm mc}\left(cm^{\dagger}+c^{\dagger}m \right) +i E_{d} \big(c^{\dagger}e^{-i\omega_{d}t}- {\rm H.c.} \big),\\
	\end{equation}
	where $c$ ($c^{\dagger}$) is the annihilation (creation) operator of the microwave cavity mode, $\omega_{c}$ is the cavity resonance frequency, and $g_{\rm mc}$ is the cavity-magnon coupling strength. Here, $E_{d}=\sqrt{2P_{d}\kappa_{c}/\hbar\omega_{d}}$ denotes the coupling strength between the cavity mode (with decay rate $\kappa_{c}$) and the microwave drive field with frequency $\omega_{d}$ and power $P_{d}$.

	\section{Preparation of non-Gaussian states}\label{III}
	
	The addition (subtraction) of a single excitation onto Gaussian states, such as coherent and thermal states~\cite{Agarwal91,Agarwal92,Sci04,PRA07} (squeezed states~\cite{Dakna,Wenger}) provides an efficient way to prepare NGSs. Here, we apply the idea to the magnonic system and propose to prepare two NGSs by adding a single magnon, in a heralded way, onto magnon coherent and thermal states, which are two most common classical states and can be readily prepared in the laboratory.  This is realized by activating an effective optomagnonic two-mode squeezing (TMS) interaction combined with the single-photon detection. 
	
\begin{figure}[t]
		\includegraphics[width=\linewidth]{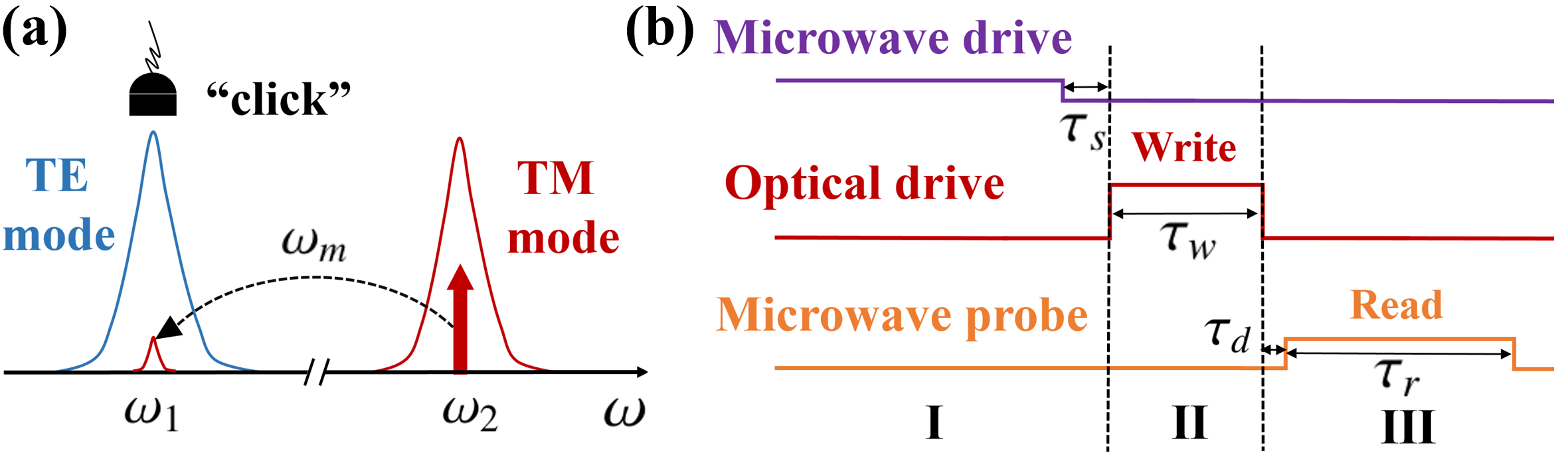}
		\caption{(a) The magnon-induced Stokes scattering used in the single-magnon addition process, where a TM-polarized photon converts into a TE-polarized photon by creating a magnon.  (b) Time sequence of the optical and microwave pulses used in the protocol, which consists of three steps. A weak microwave drive (purple line) is used to prepare an initial magnon coherent state. Once the magnon mode is in the desired state, the microwave drive is switched off. After a short period $\kappa_{c}^{-1}\ll\tau_{s}\ll\kappa_{m}^{-1}$, during which all microwave cavity photons decay, while the magnon state remains almost unchanged, an optical write pulse (red line) with duration $\tau_{w}$ is sent to activate the optomagnonic Stokes scattering. After a single photon is detected in a short interval $\tau_{d}$, indicating that the magnon mode is prepared in the desired NGS, a microwave read pulse (orange line) with duration $\tau_{r}$ is sent to the microwave cavity to map the magnon state to the cavity output field, on which subsequent measurements are performed to extract the state. To minimize the influence of dissipation on the magnon state, the total time is assumed to satisfy $\tau_{\rm total}=\tau_{s}+\tau_{w}+\tau_{d}+\tau_{r}\ll \kappa_{m}^{-1}$.} 
		\label{fig2}
	\end{figure}

	
	The optomagnonic Hamiltonian $H_{\rm OM}=H_{0}+H_{1}$ accounts for the magnon-induced BLS process, which, in the frame rotating at the drive frequency $\omega_{p_{j}}$, is given by
	\begin{equation}
		\begin{split}
			H_{\rm OM}/\hbar=&\Delta_{1}a^{\dagger}_{1}a_{1}+\Delta_{2}a^{\dagger}_{2}a_{2}+\omega_{m}m^{\dagger}m \\ &+g_{\rm om}(a_{1}^{\dagger} a_{2} m^{\dagger}+a_{1} a_{2}^{\dagger} m)
			+i E_{j}(a_{j}^{\dagger}-a_{j}),\\
		\end{split}
	\end{equation}
	where $\Delta_{j}=\omega_{j}-\omega_{p_{j}}$ ($j=1,2$) is the WGM-drive detuning. To have an optomagnonic TMS interaction, we consider the case where the WGM $a_{2}$ is resonantly pumped, i.e., $\Delta_{2}=0$ and $\Delta_{1}=-\omega_{m}$ (Fig.~\ref{fig2}(a)), realized by coupling the driving field (with a certain polarization) to the TM-polarized WGM.
	In this case, the strongly driven WGM $a_{2}$ can be treated classically as a number $\alpha_{2}\equiv\langle a_{2}\rangle=E_{2}/\kappa_{2}$ (here $\kappa_{2}$ denotes the linewidth of the WGM). This leads to the following linearized Hamiltonian in the interaction picture~\cite{Jie21X}
	\begin{equation}\label{TMSHal}
		H_{\rm TMS}/\hbar=G_{1}\big(a_{1}^{\dagger}m^{\dagger}+a_{1}m \big),
	\end{equation}
	where $G_{1}=g_{\rm om}\alpha_{2}$ is the effective optomagnonic coupling strength and accounts for the TMS interaction between the TE WGM $a_{1}$ and the magnon mode. This corresponds to the optomagnonic Stokes scattering, where a TM-polarized photon converts into a TE-polarized photon by creating a magnon excitation, as depicted in Fig.~\ref{fig2}(a).

	When the optomagnonic TMS interaction (Eq.~\eqref{TMSHal}) is sufficiently weak, the scattering giving rise to pairs of {\it single} magnons and photons is dominant, i.e., the probability of creating two-magnon/photon state $|2\rangle$ and higher excitation states is negligibly small~\cite{Simon16,Simon17,Jie18R,Jiang21}. This predicts a single magnon addition onto the magnonic initial state conditioned on the detection of a single (TE-polarized) photon.   For simplicity, we consider the drive field to be a flattop pulse with power $P_{2}$ and duration $\tau_{w}$~\cite{Hofer11,TJK14}. Assuming that the system is in an initial pure state $|\varphi(0)\rangle=|0\rangle_{1}|\psi\rangle_{m}$, the state of the system after the pulse can be approximated as (unnormalized)
	\begin{equation}
		\begin{split}
			|\varphi(\tau_{w})\rangle&=e^{-iG_{1}\big(a_{1}m+a_{1}^{\dagger}m^{\dagger}\big)\tau_{w}}|\varphi(0)\rangle\\
			&\approx|0\rangle_{1}|\psi\rangle_{m}-iG_{1}\tau_{w}|1\rangle_{1}\big(m^{\dagger}|\psi\rangle_{m} \big),\\
		\end{split}
	\end{equation}
	when $G_{1}\tau_{w}\ll 1$, 
	which indicates that the single-magnon-added state $m^{\dagger}|\psi\rangle_{m}$ can be generated on the condition that a single TE-polarized photon is detected. Note that we will derive in Sec.~\ref{thermSec} the exact solution when the magnonic initial state is not a pure state but a generic mixed state, e.g., a thermal state.

	Our protocol consists of three steps, as depicted in Fig.~\ref{fig2}(b): $i$) State initialization. To prepare a magnon coherent state, a weak microwave coherent field is used to drive the microwave cavity and the magnon mode is thereby displaced to a coherent state due to the cavity-magnon beamsplitter interaction (Eq.~\eqref{cmBS})~\cite{Jie19}.  The microwave cavity is placed at a low bath temperature to ensure that the thermal excitation of both the microwave field and the magnon mode is negligibly small and their noise is essentially vacuum noise. To prepare a magnon thermal state, we simply increase the bath temperature to have a nonzero thermal occupation. $ii$) Adding a single magnon. As explained above, a single magnon can be added onto the magnon coherent or thermal state by very weakly activating the optomagnonic Stokes scattering combined with the single-photon detection. $iii$) Readout of magnonic NGSs. The single-MACS (MATS) can be read out by sending a weak probe field to the microwave cavity and measuring the cavity output field.  This again uses the cavity-magnon state-swap interaction (Eq.~\eqref{cmBS}).

	\subsection{Magnon-added coherent states} \label{cohSec}

	In this section, we explicitly show how to generate the single-MACS.   We first prepare a magnon coherent state. By placing the system in a dilution refrigerator maintained at a temperature of, e.g., tens of mK, the magnon mode, with its resonance frequency typically in gigahertz~\cite{Nakamura14,Zhang14}, is essentially in the vacuum state with the thermal occupation  $\bar{n}_0 \approx 0$. A magnon coherent state can then be achieved by driving the microwave cavity with a coherent microwave field and the cavity-magnon beamsplitter interaction displaces the magnon mode from the vacuum state to a coherent state. 
	
	The Hamiltonian of the cavity-magnon system is $H_{\rm CM}= H_{0}+H_{2}$, which in the frame rotating at the drive frequency $\omega_{d}$ reads
	\begin{equation}
		H_{\rm CM}/\hbar= \Delta_{c}c^{\dagger}c+\Delta_{m}m^{\dagger}m+g_{\rm mc}\big(cm^{\dagger}+c^{\dagger}m \big)+iE_{d}\big(c^{\dagger}-c \big),\\
	\end{equation}
	with $\Delta_{c(m)}=\omega_{c(m)}-\omega_{d}$.  This gives rise to the following Langevin equations for the averages of the cavity and magnon modes:
	\begin{equation}
		\begin{aligned}
			&\langle\dot{c}\rangle=-(i\Delta_{c}+\kappa_{c})\langle c\rangle-ig_{\rm mc}\langle m\rangle+E_{d},\\
			&\langle\dot{m}\rangle=-(i\Delta_{m}+\kappa_{m})\langle m\rangle-ig_{\rm mc}\langle c\rangle,\\
		\end{aligned}
	\end{equation}
	with $\kappa_{m}$ being the dissipation rate of the magnon mode. For a fixed drive power and in the long-time limit, the system reaches a steady state and the corresponding solutions of the averages are
	\begin{equation}
		\begin{aligned}
			&\langle m\rangle=\dfrac{-ig_{\rm mc}E_{d}}{g_{\rm mc}^{2}-(\Delta_{c}-i\kappa_{c})(\Delta_{m}-i\kappa_{m})},\\
			&\langle c\rangle=\dfrac{(i\Delta_{m}+\kappa_{m})E_{d}}{g_{\rm mc}^{2}-(\Delta_{c}-i\kappa_{c})(\Delta_{m}-i\kappa_{m})}.
		\end{aligned}
	\end{equation}
	In the resonant case $\Delta_{c}=\Delta_{m}=0$ and under experimentally feasible parameters~\cite{Nakamura14,Zhang14,ShenNC}: $\omega_{c}/2\pi=10$ GHz, $\kappa_{c}/2\pi=40$ MHz, $\kappa_{m}/2\pi=0.5$ MHz and $g_{\rm mc}/2\pi=2$ MHz, we find that a magnon coherent state $|\beta\rangle_{m}$ with the amplitude $0.5< |\beta| \equiv |\langle m\rangle|<4$ corresponds to the microwave drive power $P_d$ ranging from 0.02 to 1.2 fW. Here, we consider a relatively large cavity decay rate to increase the drive power and to adiabatically eliminate the microwave cavity in the readout process in Sec.~\ref{IV}. The drive power can also be increased by taking a nonzero detuning $\Delta_{c(m)}$. Note that $\langle m\rangle$ is generally complex, but can be set to be real by adjusting the phase of the drive field.

	Once the magnon mode is prepared in a desired coherent state, we switch off the microwave drive, and after a short time $\kappa_{c}^{-1}\ll\tau_{s}\ll\kappa_{m}^{-1}$, during which all microwave cavity photons dissipate, while the magnon state remains almost unchanged, we then send an optical write pulse (with duration $\tau_{w}$, Fig.~\ref{fig2}(b)) to drive the TM-polarized WGM to activate the optomagnonic Stokes scattering. Assuming that $\tau_{w}\ll\kappa_{m}^{-1}$, the dissipation of the magnon mode during the write pulse can be neglected~\cite{Jie18R,TJK14}. This leads to the following quantum Langevin equations (QLEs) during the pulse
	\begin{equation}
		\begin{aligned}
			&\dot{a}_{1}=-\kappa_{1}a_{1}-iG_{1}m^{\dagger}+\sqrt{2\kappa_{1}}a_{1}^{\rm in},\\
			&\dot{m}=-iG_{1}a_{1}^{\dagger},\\
		\end{aligned}
	\end{equation}
	where $\kappa_{1}$ ($a_{1}^{\rm in}$) is the decay rate (input field) of the WGM. We consider a weak coupling $G_{1}\ll\kappa_{1}$, which allows for the adiabatic elimination of the WGM, yielding $a_{1}\simeq\kappa_{1}^{-1}(-iG_{1}m^{\dagger}+\sqrt{2\kappa_{1}}a_{1}^{\rm in})$. By further using the input-output relation $a_{1}^{\rm out}=\sqrt{2\kappa_{1}}a_{1}-a_{1}^{\rm in}$, we get 
	\begin{equation}\label{a1-m}
		\begin{aligned}
			&a_{1}^{\rm out}=-i\sqrt{2\mathcal{G}_{1}}m^{\dagger}+a_{1}^{\rm in},\\
			&\dot{m}=\mathcal{G}_{1}m-i\sqrt{2\mathcal{G}_{1}}a_{1}^{\rm in \dagger},\\
		\end{aligned}
	\end{equation}
	with $\mathcal{G}_{1}\equiv G_{1}^{2}/\kappa_{1}$. We further define a set of normalized temporal modes~\cite{Hofer11} for the WGM driven by a pulse with duration $\tau_{w}$
	\begin{equation}
		\begin{aligned}
			&A_{1}^{\rm in}(\tau_{w})=\sqrt{\dfrac{2\mathcal{G}_{1}}{1-e^{-2\mathcal{G}_{1}\tau_{w}}}}\int_{0}^{\tau_{w}}e^{-\mathcal{G}_{1}s}a_{1}^{\rm in}(s)ds,\\
			&A_{1}^{\rm out}(\tau_{w})=\sqrt{\dfrac{2\mathcal{G}_{1}}{e^{2\mathcal{G}_{1}\tau_{w}}-1}}\int_{0}^{\tau_{w}}e^{\mathcal{G}_{1}s}a_{1}^{\rm out}(s)ds,\\
		\end{aligned}
	\end{equation}
	which satisfy the canonical commutation relation $[A_{1}^{k},A_{1}^{k\dagger}]=1$ ($k=$in, out). By integrating Eq.~\eqref{a1-m}, we obtain
	\begin{equation}\label{aaaaa}
		\begin{aligned}
			&A_{1}^{\rm out}(\tau_{w})=-i\sqrt{e^{2\mathcal{G}_{1}\tau_{w}}-1}m^{\dagger}(0)+e^{\mathcal{G}_{1}\tau_{w}}A_{1}^{\rm in}(\tau_{w}),\\
			&m(\tau_{w})=e^{\mathcal{G}_{1}\tau_{w}}m(0)-i\sqrt{e^{2\mathcal{G}_{1}\tau_{w}}-1}A_{1}^{\rm in\dagger}(\tau_{w}).
		\end{aligned}
	\end{equation}
	Equation~\eqref{aaaaa} allows us to extract a propagator $U(\tau_{w})$ satisfying $A_{1}^{\rm out}(\tau_{w})=U^{\dagger}(\tau_{w})A_{1}^{\rm in}(\tau_{w})U(\tau_{w})$ and $m({\tau_w})=U^{\dagger}(\tau_{w})m(0)U(\tau_{w})$, given by~\cite{TJK14}
	\begin{equation}\label{eq:13}
		U(\tau_{w})=e^{-i\sqrt{1-M^{2}(\tau_{w})}A_{1}^{\rm in\dagger}m^{\dagger}}M(\tau_{w})^{1+A_{1}^{\rm in\dagger}A_{1}^{\rm in}+m^{\dagger}m}e^{i\sqrt{1-M^{2}(\tau_{w})}A_{1}^{\rm in}m},
	\end{equation}
	where $M(\tau_{w})=e^{-\mathcal{G}_{1}\tau_{w}}$ ($0<M < 1$). 
	For an initial state of the optomagnonic system $|0\rangle_{1}|\beta\rangle_{m}$, the system, at the end of the write pulse, is prepared in the state (unnormalized)
	\begin{equation}\label{bbbb}
		\begin{split}
			|\varphi(\tau_{w})\rangle
			&=\sum_{n=0}^{\infty}\dfrac{i^{n}\big(1-M^{2}\big)^{\frac{n}{2}}}{n!}\big(A_{1}^{\rm in\dagger}m^{\dagger} \big)^{n}|0\rangle_{1}|M\beta\rangle_{m}\\
			&\approx|0\rangle_{1}|M\beta\rangle_{m}-i\sqrt{1-M^{2}}|1\rangle_{1}\big(m^{\dagger}|M\beta\rangle_{m}\big).
		\end{split}
	\end{equation}
	We have omitted higher-order terms ($n \ge 2$), which is a good approximation when $1-M^{2}\ll 1$, i.e., $\mathcal{G}_{1}\tau_{w}\ll 1$. It indicates that the single-MACS $m^{\dagger}|M\beta\rangle_{m}$ is generated when a single TE-polarized photon is detected and the heralding probability is approximately $1-M^{2}$.  Note that the amplitude of the magnon coherent state is slightly reduced, since $|\beta|\rightarrow M |\beta| = |\beta|e^{-\mathcal{G}_{1}\tau_{w}}$.

	\begin{figure}[b]
		\hskip-0.4cm \includegraphics[width=0.82\linewidth]{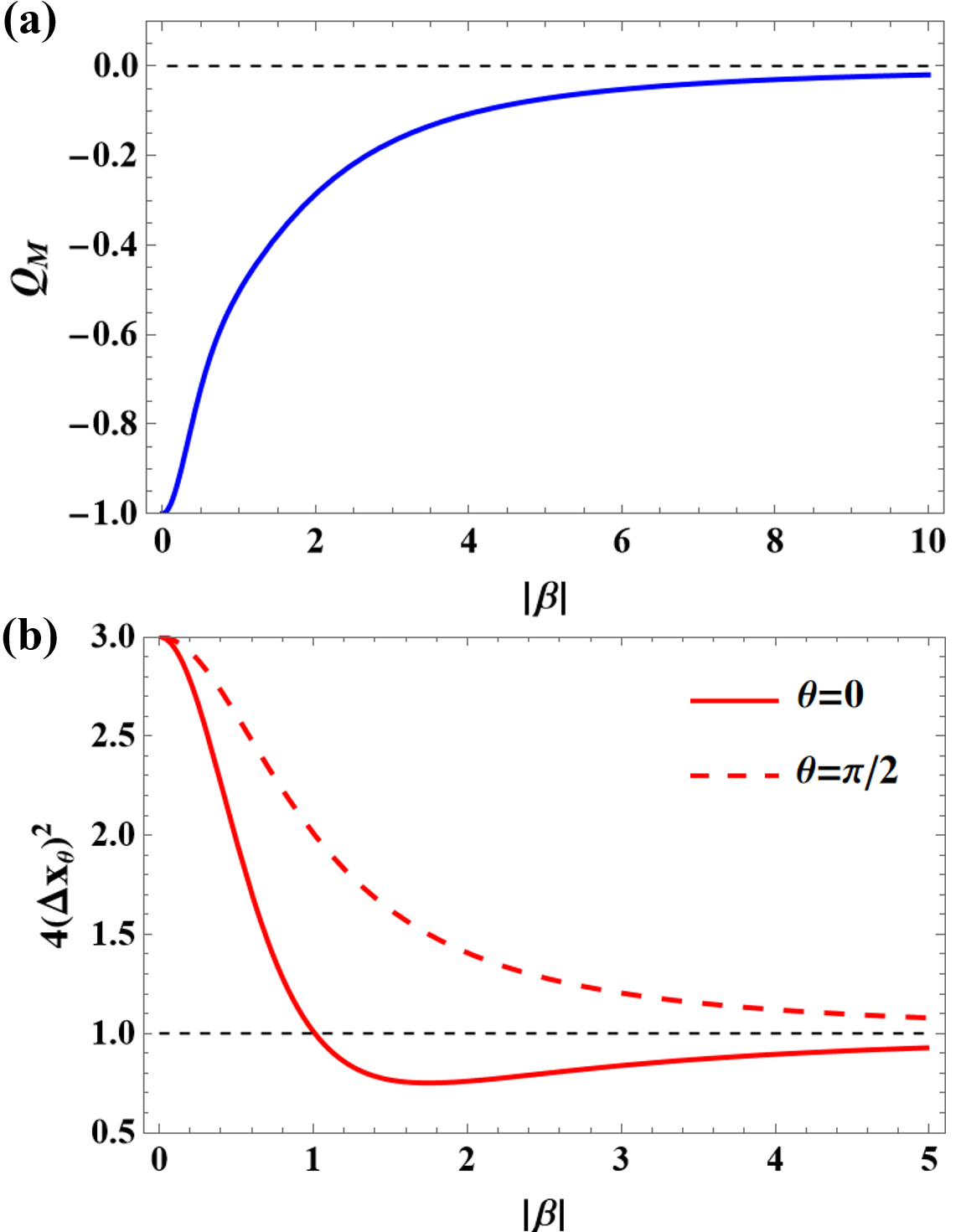}
		\caption{(a) Mandel $Q$ parameter versus the amplitude $|\beta|$ of the initial coherent state. (b) Variance $4(\Delta x_{\theta})^{2}$ as a function of $|\beta|$ for $\theta=0$ (solid line) and $\theta=\pi/2$ (dashed line). The horizontal dashed line denotes $Q_{M}=0$ in (a) or $4(\Delta x_{\theta})^{2}=1$ in (b) for the coherent state. We use promising parameters $\tau_{w}=16$ ns $\ll\kappa_{m}^{-1}$, $G_{1}/2\pi=1$ MHz, and $\kappa_{1}/2\pi=10$ MHz, which yield $\mathcal{G}_{1}\tau_{w}\approx 0.01$.}
		\label{fig3}
	\end{figure}

	The above single-MACS is a NGS, which exhibits many nonclassical properties, including the sub-Poissonian character of the magnon-number distribution, quadrature squeezing, and  a negative Wigner function. We first study the magnon-number distribution of the state by calculating the Mandel $Q$ parameter, which is defined as $Q_{M}=\big(\langle m^{\dagger 2}m^{2}\rangle-\langle m^{\dagger}m\rangle^{2} \big)/\langle m^{\dagger}m\rangle$~\cite{Mandel}. A negative $Q_{M}<0$ represents the sub-Poissonian character of the state and is a signature of non-classicality~\cite{Agarwal91,Agarwal92}, since classical states, such as coherent and thermal states, correspond to Poissonian ($Q_{M}=0$) or super-Poissonian ($Q_{M}>0$) statistics.  After some calculation, we obtain $Q_{M}$ of the single-MACS, i.e.,
	\begin{equation}\label{eq:15}
		Q_{M}=-\dfrac{1+2M^{2}|\beta|^{2}+2M^{4}|\beta|^{4}}{1+M^{2}|\beta|^{2}(2+M^{2}|\beta|^{2})^{2}}.
	\end{equation}
	In Fig.~\ref{fig3}(a), we show the $Q$ parameter versus the amplitude $|\beta|$ of the initial magnon coherent state. Clearly, a negative $Q_{M}$ is present and a small amplitude $|\beta|$ is preferred for seeing a strong sub-Poissonian feature.  Note that $|\beta|=0$ corresponds to the single-magnon state $|1\rangle_m$, of which the $Q$ parameter $Q_{M}=-1$, and for a large amplitude $|\beta|\to\infty$, $Q_{M}\to 0$, implying that the state tends to be a coherent state.  Therefore, the single-MACS $m^{\dagger}|M\beta\rangle_{m}$ is in an intermediate state between the single-magnon state (a fully quantum state) and the coherent state (a classical state), which can be exploited to study the quantum-to-classical transition~\cite{Sci04}.

	Another nonclassical feature of the single-MACS is the quadrature squeezing~\cite{Agarwal91}, which both coherent and Fock states do not possess. To uncover this, we define a general quadrature of the magnon mode $x_{\theta}=\frac{1}{2}\big(me^{i\theta}+m^{\dagger}e^{-i\theta}\big)$. It is squeezed if its variance $\big(\Delta x_{\theta}\big)^{2}=\langle x_{\theta}^{2}\rangle-\langle x_{\theta}\rangle^{2}$ is smaller than that of the vacuum state. The expression of the variance $(\Delta x_{\theta})^{2}$ is given by
	\begin{equation}
		\big(\Delta x_{\theta}\big)^{2}=\dfrac{1-M^{2}|\beta|^{2}\cos 2\theta}{2(1+M^{2}|\beta|^{2})^{2}}+\dfrac{1}{4}.
	\end{equation}
	In Fig.~\ref{fig3}(b), we plot $4(\Delta x_{\theta})^{2}$ versus the amplitude of the coherent state $|\beta|$, and consider two cases: $\theta=0$ and $\frac{\pi}{2}$, corresponding to the optimal $\theta$ for squeezing and anti-squeezing, respectively. In our definition, $4(\Delta x_{\theta})^{2}=1$ corresponds to the vacuum fluctuation. The initial coherent state has equal vacuum fluctuation in different quadratures independently from the amplitude, while the single-MACS exhibits squeezing in one quadrature ($\theta=0$) but anti-squeezing in the orthogonal quadrature ($\theta=\frac{\pi}{2}$) when $|\beta|>1$.  

	\begin{figure}[t]
		\includegraphics[width=\linewidth]{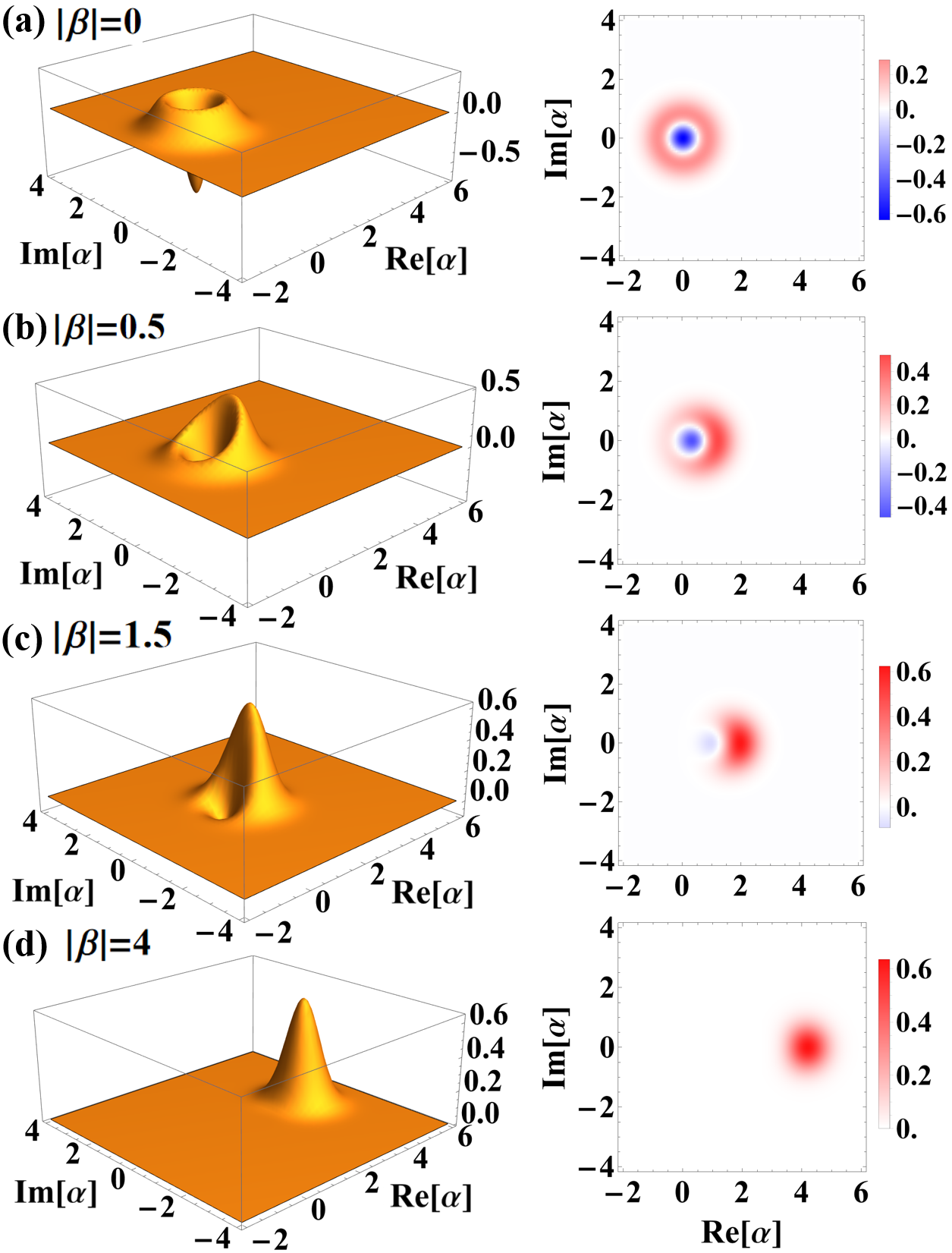}
		\caption{Wigner function of the single-MACS for different values of $|\beta|$: (a) $|\beta|=0$, (b) $|\beta|=0.5$, (c) $|\beta|=1.5$, and (d) $|\beta|=4$. The other parameters are the same as in Fig.~\ref{fig3}.}
		\label{fig4}
	\end{figure}

	\begin{figure}[b]
		\hskip-0.5cm \includegraphics[width=0.92\linewidth]{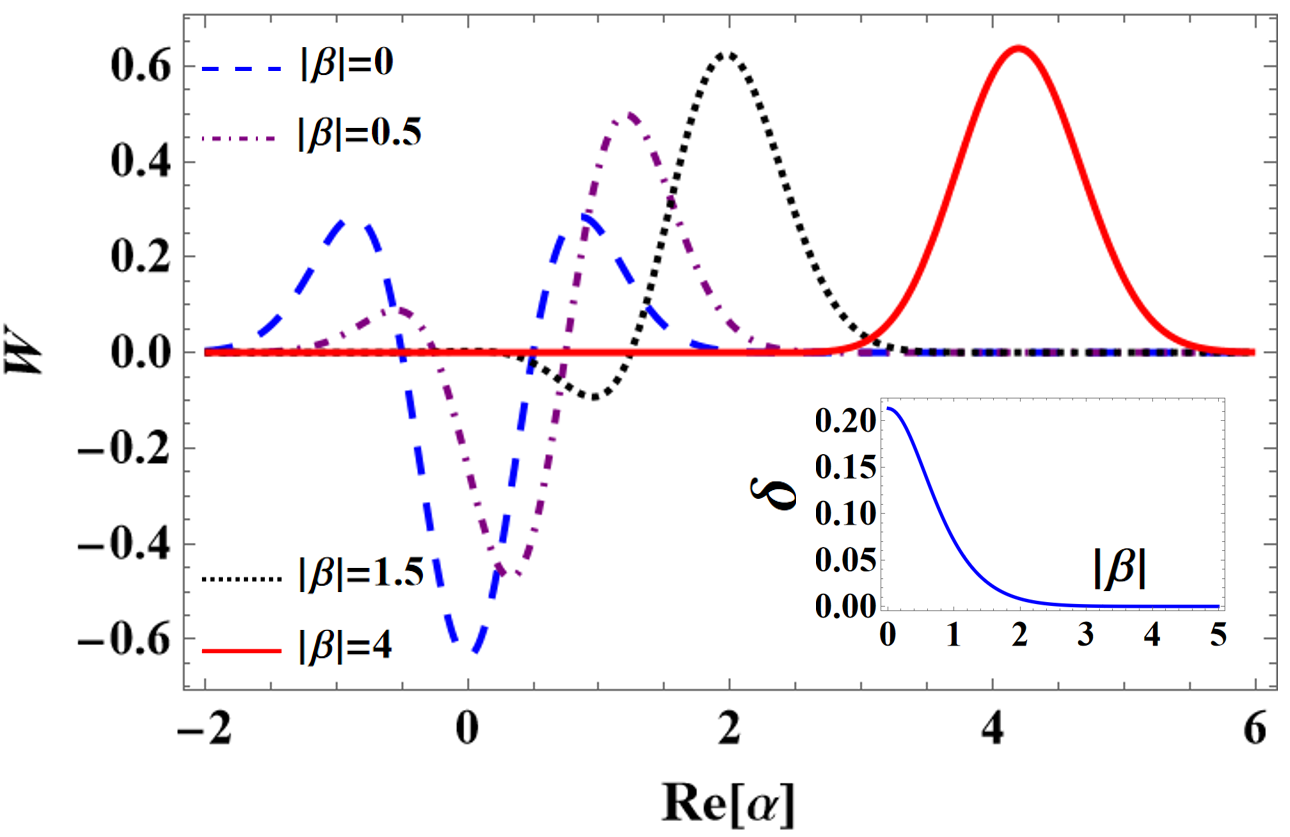}
		\caption{Wigner function [the Im$(\alpha)=0$ plane] of the single-MACS for different values of $|\beta|$. Inset: Wigner negativity $\delta$ as a function of $|\beta|$. The other parameters are those as in Fig.~\ref{fig3}.}
		\label{fig5}
	\end{figure}

	We further calculate the Wigner function~\cite{Wigner} of the single-MACS $m^{\dagger}|M\beta\rangle_{m}$, which reads
	\begin{equation}
		\begin{split}
			W(\alpha)=&\dfrac{2}{\pi \left(1+M^{2}|\beta|^{2} \right)} e^{-2\big[|\alpha|^{2}+M^{2}|\beta|^{2}-M|\beta|(\alpha+\alpha^{*}) \big]}  \\
			&\times \left[4|\alpha|^{2}-2M|\beta|\big(\alpha+\alpha^{*} \big)+M^{2}|\beta|^{2}-1 \right], \\
		\end{split}
	\end{equation}
	where $\alpha$ is a complex variable. It is known that the negativity of the Wigner function is an indicator of both non-classicality and non-Gaussianity. For instance, a coherent state manifests as a Gaussian packet without any negativity in the phase space, whereas a Fock state displays negative values.   In Fig.~\ref{fig4}, we show the Wigner function for the single-MACS with different values of $|\beta|$. In the limit of $|\beta|=0$, the state becomes the single-magnon state which exhibits negative values around the origin of the phase space (Fig.~\ref{fig4}(a)).  By increasing $|\beta|$, the negativity is gradually lost (Fig.~\ref{fig4}(b)-(c)) and the Wigner function eventually tends to be that of a coherent state (Fig.~\ref{fig4}(d)). This process corresponds to a smooth transition from a quantum state to a classical state, as demonstrated in photonic systems~\cite{Sci04,PRA05}.
	This is more clearly seen in the slice of the Wigner function as shown in Fig.~\ref{fig5}. The inset shows a continuous decrease of the Wigner negativity (defined as the volume $\delta$ of negative Wigner distributions in the phase space~\cite{Kenfack}) as a function of $|\beta|$. Clearly, the state gradually loses its non-classicality and non-Gaussianity as $|\beta|$ increases.



	\subsection{Magnon-added thermal states} \label{thermSec}

	As the most common classical state, the thermal state can become a non-Gaussian and nonclassical state by adding a single excitation~\cite{Agarwal92}. For a magnon mode of a YIG sphere, this turns to be a single-MATS, which is a macroscopic quantum state and shows counterintuitive features, e.g., the mean thermal excitation number is doubled by adding a single excitation, as demonstrated in phononic systems~\cite{Enzian21,Patel21}. Moreover, it exhibits nonclassical characteristics that differ from those of the single-MACS, and thus opens up new avenues for  preparing nonclassical states with the minimum requirement for the initial state.
	
	An initial thermal state of the magnon mode can be achieved by simply operating the system at a bath temperature giving a nonzero thermal occupation $\bar{n}_{0}>0$. For the magnon frequency of, e.g., $\omega_{m}/2\pi=10$ GHz, the temperature $0.2 \,{\rm K} <T<1$ K corresponds to a magnon thermal state with $0.417<\bar{n}_{0}<2.084$. The thermal state is expressed as
	\begin{equation}
		\rho_{\rm th}=(1-p)\sum_{n=0}p^{n}|n\rangle\langle n|,\quad \,\, p\equiv\dfrac{\bar{n}_{0}}{1+\bar{n}_{0}}.
	\end{equation}
	Similarly as in Sec.~\ref{cohSec}, we send an optical write pulse to the TM-polarized WGM to implement the operation of the single-magnon addition. Assuming that the optomagnonic system is in an initial state $|0\rangle_{1}\langle 0|_{1}\otimes\rho_{\rm th}$ and using the propagator $U(\tau_{w})$ defined in Eq.~\eqref{eq:13}, the system, at the end of the pulse, is prepared in the following mixed state (unnormalized)
	\begin{equation}\label{ccccc}
		\rho(\tau_{w})
		\approx|0\rangle_{1}\langle 0|_{1}\otimes \Big[  \rho'_{\rm th} \Big]_{m}+\big(1-M^{2} \big)|1\rangle_{1}\langle 1|_{1} \otimes \Big[ m^\dag \rho'_{\rm th} m \Big]_{m},\\		
	\end{equation}
	where the thermal state
	\begin{equation}
		\rho'_{\rm th} = (1-p')\sum_{n=0}(p')^{n}|n\rangle\langle n|, \quad \,\,  p'=pM^{2} \equiv  \dfrac{\bar{n}}{1+\bar{n}}, 
	\end{equation}
	with a reduced thermal occupation due to the write pulse
	\begin{equation}
		\bar{n}=\dfrac{\bar{n}_{0}M^{2}}{1+(1-M^{2})\bar{n}_{0}} < \bar{n}_{0}.
	\end{equation}
	Equation~\eqref{ccccc} indicates that the magnon mode is prepared in a single-MATS, i.e., $m^\dag \rho'_{\rm th} m =  (1-p')\sum_{n=0}(p')^{n}(n+1)|n+1\rangle\langle n+1|$ (unnormalized), if a single TE-polarized photon is detected. When no photon is created,  the magnon mode remains in the thermal state  $\rho'_{\rm th}$, but with a reduced thermal occupation $\bar{n}_{0} \to \bar{n}$.

	\begin{figure}[t]
		\hskip-0.3cm \includegraphics[width=0.85\linewidth]{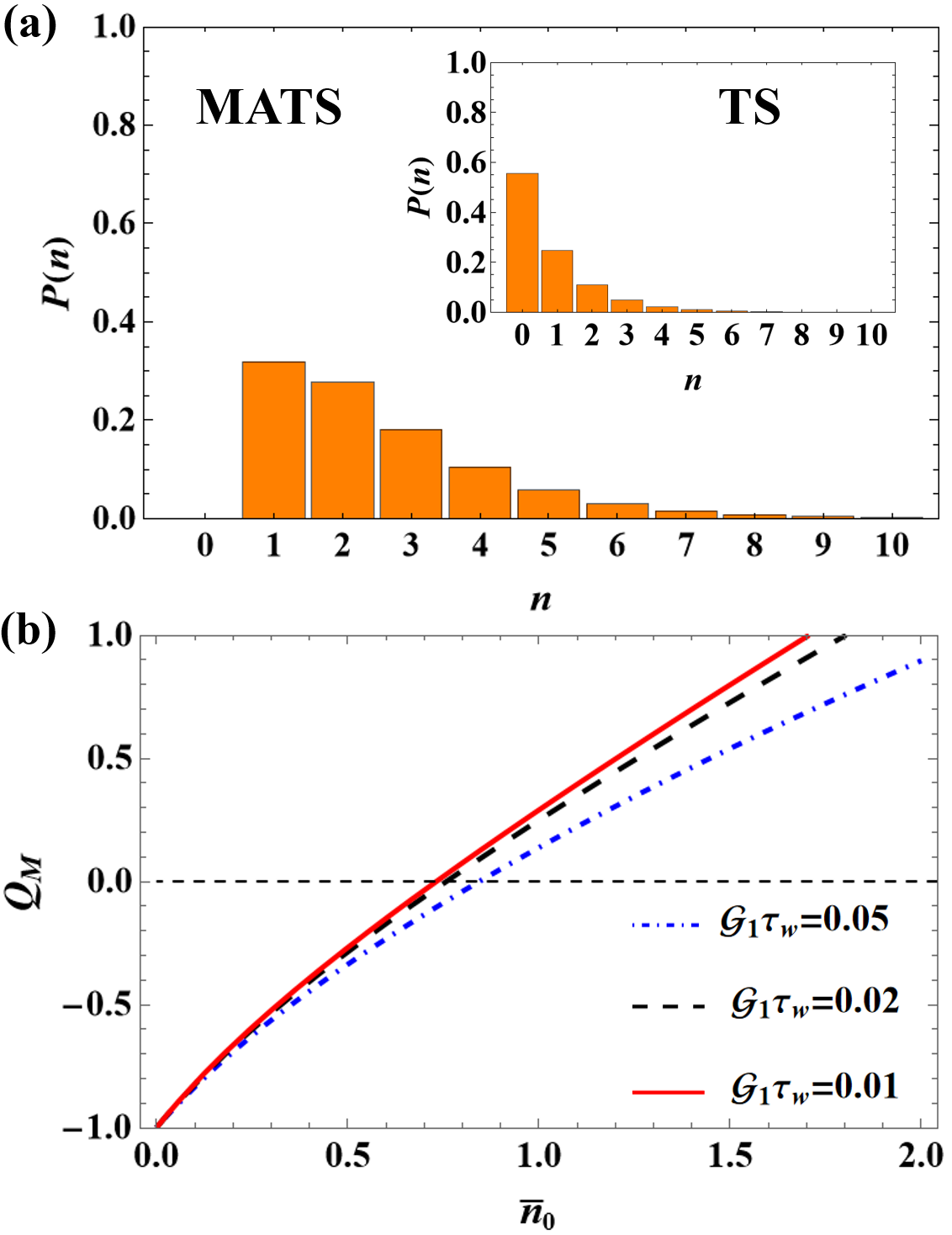}
		\caption{(a) Magnon-number distribution of the single-MATS and thermal state (inset). We take the initial thermal occupation $\bar{n}_{0}=0.8$ and $\mathcal{G}_{1}\tau_{w}=0.01$ as in Fig.~\ref{fig3}. (b) Mandel parameter $Q_{M}$ versus $\bar{n}_{0}$ for different values of $\mathcal{G}_{1} \tau_{w}$. We take $\tau_{w}=16$ ns, and $\mathcal{G}_1$ is altered by changing the power of the write pulse and $0.01<\mathcal{G}_{1}\tau_{w}<0.05$ corresponds to the effective optomagnonic coupling $1$ MHz $<G_{1}/2\pi<2.23$ MHz. The other parameters are the same as in Fig.~\ref{fig3}. }
		\label{fig6}
	\end{figure}

	Though having no squeezing, the single-MATS can exhibit the sub-Poissonian feature in the magnon-number distribution indicated by a negative $Q$ parameter~\cite{Agarwal92}.  
	In Fig.~\ref{fig6}(a), we plot the magnon-number distributions of the thermal state and single-MATS. It shows that by adding a single magnon, the number distribution of the state is changed significantly. This manifests as a vacancy in the vacuum state, {which has the highest probability in the initial thermal state but with zero contribution to the mean excitation number}, while the remaining number states are renormalized~\cite{PRA07}. The renormalized number distribution causes a counterintuitive effect, i.e., the thermal occupation is approximately doubled by adding a single excitation, $\bar{n}\rightarrow 2\bar{n}+1$~\cite{Enzian21,Patel21}.  
	We obtain the Mandel $Q$ parameter of the single-MATS, given by
	\begin{equation}
		Q_{M}=\dfrac{2M^{2}\bar{n}_{0}(\bar{n}_{0}+1)}{[(1+M^{2})\bar{n}_{0}+1][(1-M^{2})\bar{n}_{0}+1]}-1.
	\end{equation}
	In Fig.~\ref{fig6}(b), we show $Q_{M}$ versus the mean thermal magnon number $\bar{n}_{0}$ for different values of $\mathcal{G}_{1}\tau_{w}$. Clearly, a small $\bar{n}_{0}$ yields a negative $Q_{M}<0$, signifying the sub-Poissonian character of the state. In fact, the condition for the sub-Poissonian statistics in our case is~\cite{Agarwal92}
	\begin{equation}
		\bar{n}=\dfrac{\bar{n}_{0}M^{2}}{1+(1-M^{2})\bar{n}_{0}} < \Big(\frac{N}{N+1} \Big)^{\frac{1}{2}},
	\end{equation}
	where $N$ represents the number of magnons added onto the thermal state and $N=1$ in our scheme.  As $\bar{n}_{0}$ increases, the state exhibits a transition from the sub-Poissonian ($Q_{M}<0$) to the super-Poissonian ($Q_{M}>0$) statistics, which can be interpreted as a quantum-to-classical transition, similar to the single-MACS by varying $|\beta|$ as  discussed in Sec.~\ref{cohSec}.

\begin{figure}[t]
		\hskip-0.5cm \includegraphics[width=0.93\linewidth]{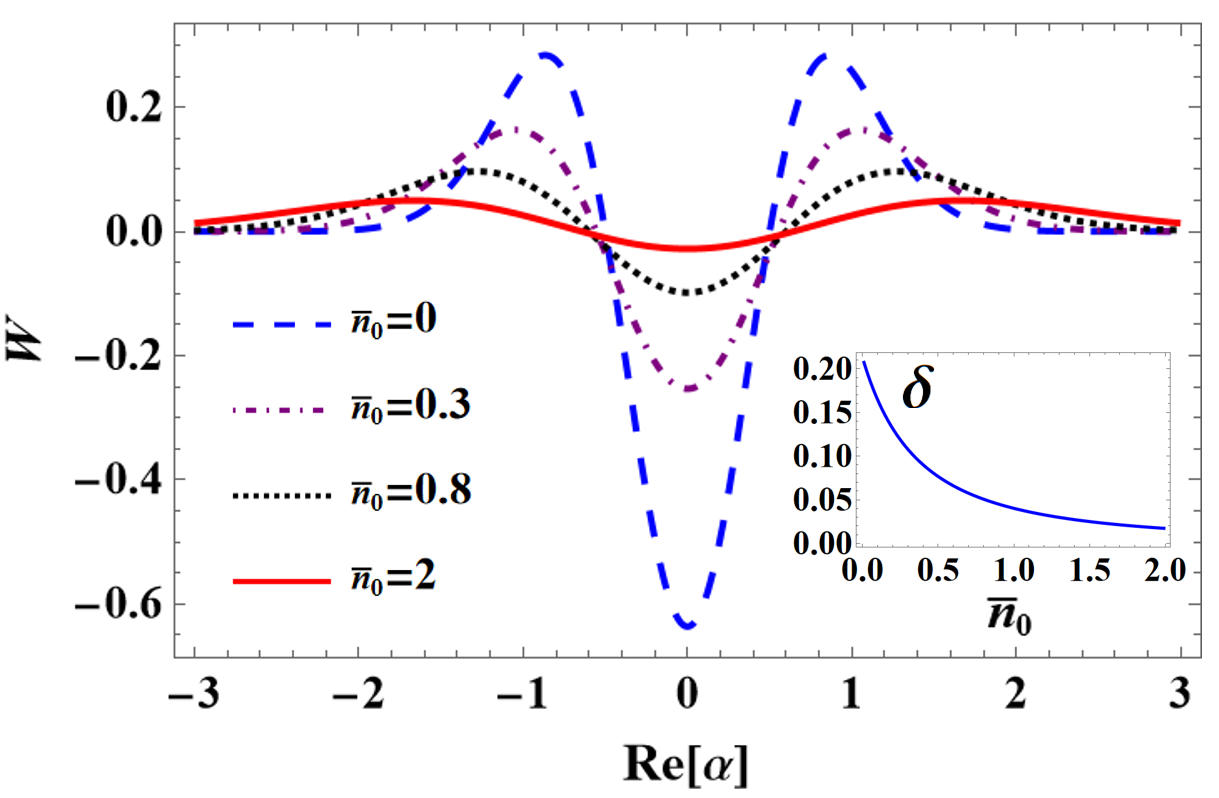}
		\caption{Wigner function [the Im$(\alpha)=0$ plane] of the single-MATS for various initial thermal occupation $\bar{n}_{0}$. Inset shows the Wigner negativity $\delta$ versus $\bar{n}_{0}$. The other parameters are the same as in Fig.~\ref{fig3}.}
		\label{fig7}
	\end{figure}

	We also calculate the Wigner function of the single-MATS, which reads
	\begin{equation}
		\begin{split}
			W(\alpha)=&\dfrac{2\left(1-\bar{n}_{0}+M^{2}\bar{n}_{0} \right)^{2}}{\pi \left(1+\bar{n}_{0}+M^{2}\bar{n}_{0} \right)^{3}} e^{-2|\alpha|^{2}\frac{1+ \bar{n}_{0} -M^{2} \bar{n}_{0}}{1+\bar{n}_{0}+M^{2}\bar{n}_{0}}} \\
			&\times\left[\left(4|\alpha|^{2}-1 \right)(\bar{n}_{0}+1)-M^{2}\bar{n}_{0} \right].\\
		\end{split}
	\end{equation}
	In Fig.~\ref{fig7}, we show the slice of the Wigner function of the single-MATS for various initial thermal occupation $\bar{n}_{0}$.  In the limit of $\bar{n}_{0}=0$, the single-MATS becomes the very nonclassical single-magnon state $|1\rangle_m$, which shows a negative Wigner distribution around the origin of the phase space. By increasing $\bar{n}_{0}$ (via rising the bath temperature), the Wigner function gradually loses its negativity, as shown in the inset.   
	It is worth noting that the single-MATS exhibits non-Gaussianity even for a large $\bar{n}_{0}$~\cite{Patel21,Enzian21}. This is counterintuitive, since for a large thermal occupation $\bar{n}_{0}$, the Wigner function would be expected to closely resemble a Gaussian shape of a thermal state.

	\section{Readout of non-Gaussian states}\label{IV}
	
	In the preceding section, we show that the magnon mode of the YIG sphere can be prepared in a single-MACS (MATS). To verify these non-Gaussian and nonclassical states, we utilize the microwave cavity and the magnon-cavity state-swap interaction, which allows us to map the magnon state to the cavity output field from which the quantum state is measured.

	Specifically, we send a microwave probe field, simplified as a flattop pulse with duration $\tau_{r}$, to the cavity (cf. Fig.~\ref{fig2}(b)), and the Hamiltonian of the cavity-magnon system is $H_{0}+H_{2}$.  The beamsplitter interaction $\hbar g_{\rm mc}(cm^{\dagger}+c^{\dagger}m)$ can coherently map the magnon state to the cavity output field~\cite{Jie18R}, as will be shown later. Again, we consider a short read pulse with duration $\tau_{r}\ll\kappa_{m}^{-1}$ to neglect the magnon dissipation during the pulse. For the resonant case $\Delta_{c}=\Delta_{m}=0$, the corresponding QLEs, in the frame rotating at the probe frequency, read
	\begin{equation}
		\begin{aligned}
			&\dot{c}=-\kappa_{c}c-ig_{\rm mc}m+\sqrt{2\kappa_{c}}c^{\rm in},\\
			&\dot{m}=-ig_{\rm mc}c,\\
		\end{aligned}
	\end{equation}
	with $c^{\rm in}$ being the input noise of the microwave cavity. Under the condition of a weak coupling $g_{\rm mc}\ll\kappa_{c}$, we can adiabatically eliminate the microwave cavity and obtain $c \simeq \kappa_{c}^{-1} (-ig_{\rm mc}m+\sqrt{2\kappa_{c}}c^{\rm in})$.  By further using the input-output relation $c^{\rm out}=\sqrt{2\kappa_{c}}c-c^{\rm in}$, introducing the temporal modes
	\begin{equation}
		\begin{aligned}
			&C^{\rm in}(\tau_{r})=i\sqrt{\dfrac{2\mathcal{G}_{c}}{e^{2\mathcal{G}_{c}\tau_{r}-1}}}\int_{0}^{\tau_{r}}e^{\mathcal{G}_{c}s}c^{\rm in}(s)ds,\\
			&C^{\rm out}(\tau_{r})=i\sqrt{\dfrac{2\mathcal{G}_{c}}{1-e^{-2\mathcal{G}_{c}\tau_{r}}}}\int_{0}^{\tau_{r}}e^{-\mathcal{G}_{c}s}c^{\rm out}(s)ds,\\
		\end{aligned}
	\end{equation}
	with $\mathcal{G}_{c} \equiv g_{\rm mc}^2/\kappa_c$, and following the same procedures as in Sec.~\ref{cohSec}, we achieve~\cite{Jie21X}
	\begin{equation}
		C^{\rm out}(\tau_{r})=\sqrt{\eta(\tau_{r})}m(0)+\sqrt{1-\eta(\tau_{r})}C^{\rm in}(\tau_{r}),
	\end{equation}
	where $\eta(\tau_{r})=1-e^{-2\mathcal{G}_{c}\tau_{r}}$ denotes the state-swap efficiency of the read pulse. Clearly, $C^{\rm out}(\tau_{r})=m(0)$ when $\eta\to1$, which implies that the magnon state is perfectly mapped to the cavity output field. Therefore, the generated magnonic NGS can be verified by measuring the cavity output field of the read pulse.

	\begin{figure}[t]
		\hskip-0.4cm	\includegraphics[width=0.8\linewidth]{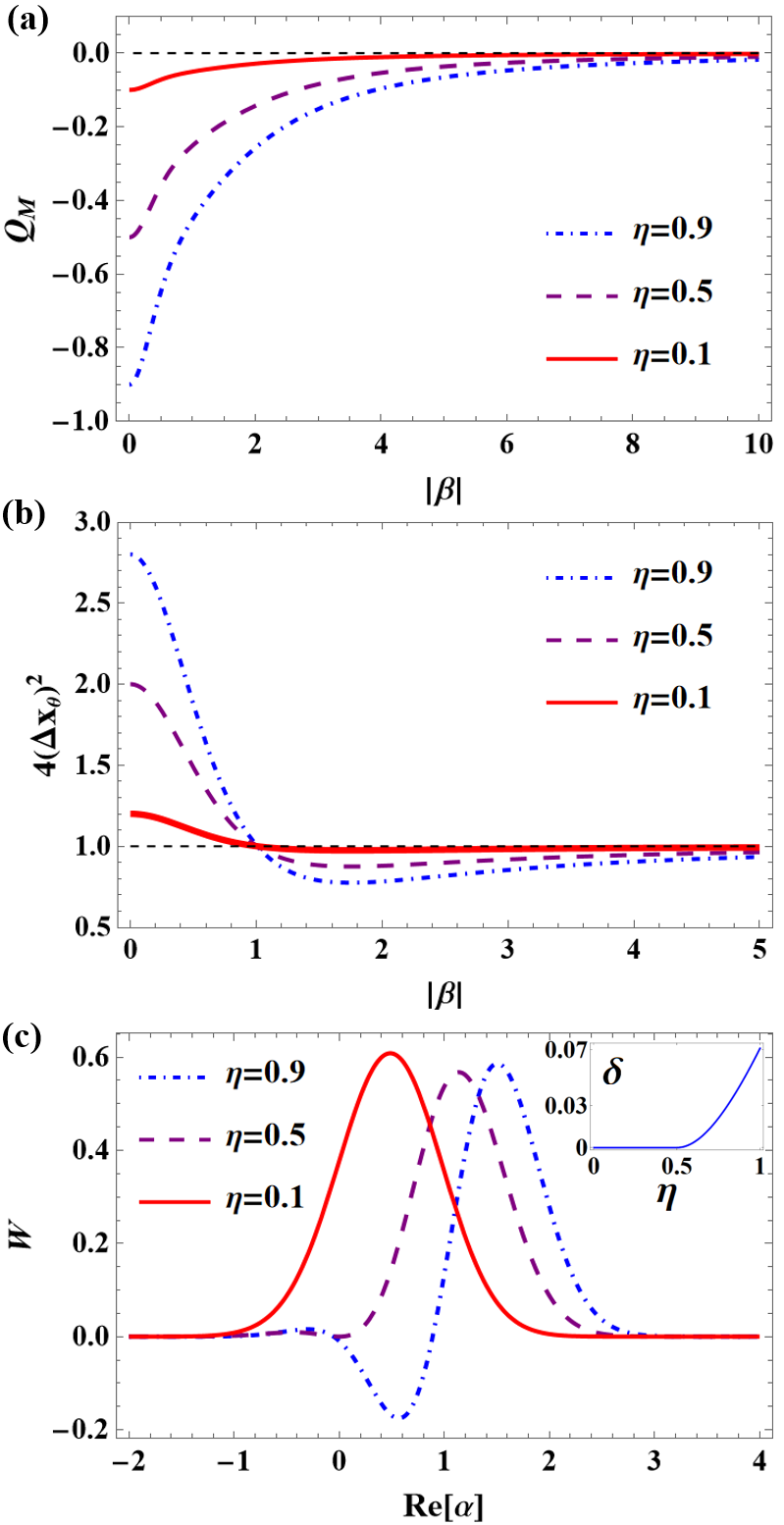}
		\caption{(a) Mandel $Q$ parameter and (b) $4(\Delta x_{\theta})^{2}$ at $\theta=0$ versus $|\beta|$ for various state-swap efficiency $\eta$. The horizontal dashed line denotes $Q_{M}=0$ or $4(\Delta x_{\theta})^{2}=1$ for the coherent state. (c) Wigner function [the Im$(\alpha)=0$ plane] of the microwave output field for various $\eta$. We take $|\beta|=1$. Inset shows the Wigner negativity $\delta$ versus the efficiency $\eta$. We use a relatively longer pulse $\tau_{r}=70$ ns to relax the constraint on the weak coupling $g_{\rm mc}\ll\kappa_{c}=2\pi \times 40$ MHz, and meanwhile to have a high efficiency $\eta$. Under these parameters, $0.1<\eta<0.9$ corresponds to $g_{\rm mc}/2\pi$ ranging from 2.2 to 10.2 MHz. The other parameters are the same as in Fig.~\ref{fig3}. }
		\label{fig8}
	\end{figure}

	Firstly, we consider the case of the single-MACS $m^{\dagger}|M\beta\rangle_{m}$. To see its sub-Poissonian character, we perform the photon-number detection of the microwave output field to calculate its Mandel $Q$ parameter. 
	Assuming an initial vacuum state of the microwave cavity, we obtain
	\begin{equation}
		Q_{M}=-\dfrac{\eta(1+2M^{2}|\beta|^{2}+2M^{4}|\beta|^{4})}{1+M^{2}|\beta|^{2}(2+M^{2}|\beta|^{2})^{2}}.
	\end{equation}
	In Fig.~\ref{fig8}(a), we show $Q_{M}$ versus the amplitude $|\beta|$ of the coherent state for various state-swap efficiency $\eta$. Obviously, the sub-Poissonian character ($Q_{M}<0$) of the magnon state is mapped to the cavity output field when $\eta$ is not too small, implying that the $Q$ parameter is robust against the noise that enters the readout process.
	
	The squeezing of the single-MACS can be verified by measuring the quadrature of the cavity output field, $x_{\theta}=\frac{1}{2}\big(C^{\rm out}e^{i\theta}+C^{\rm out \dagger}e^{-i\theta}\big)$, and its variance is given by
	\begin{equation}
		(\Delta x_{\theta})^{2}=\dfrac{\eta(1-M^{2}|\beta|^{2}\cos 2\theta)}{2(1+M^{2}|\beta|^{2})^{2}}+\dfrac{1}{4},
	\end{equation}
	which is shown in Fig.~\ref{fig8}(b) for various $\eta$. It indicates that a high state-swap efficiency $\eta$ is required to efficiently transfer the relatively small squeezing from the magnon mode to the microwave field.
	
	We further calculate the Wigner function of the cavity output field, i.e., 
	\begin{equation}
		\begin{split}						      
			W(\alpha)=&\left\{ 4\eta|\alpha|^{2} {-} (2\eta {-} 1)\left[ 4 \!\! \sqrt{\eta}M|\beta| {\rm Re}(\alpha) {-} M^{2}|\beta|^{2}(2\eta {-}1)\,{+} 1 \right] \right\}\\
			& \times\dfrac{2}{\pi \left(1+M^{2}|\beta|^{2}\right)} e^{-2\left [|\alpha|^{2}- 2\sqrt{\eta}M|\beta| {\rm Re}(\alpha)+\eta M^{2}|\beta|^{2}\right]},\\
		\end{split}
	\end{equation}
	which can be reconstructed by implementing a microwave homodyne tomography~\cite{Mallet11}.
	Figure~\ref{fig8}(c) shows the slice of the Wigner function of the output field for $|\beta|=1$ and various $\eta$. 
	It tells that a higher state-swap efficiency $\eta>0.5$ is required~\cite{PRA05,PRA07} in order to have a negative Wigner function of the output field, as more clearly seen in the inset of Fig.~\ref{fig8}(c).
	

	\begin{figure}[t]
		\hskip-0.4cm	\includegraphics[width=0.8\linewidth]{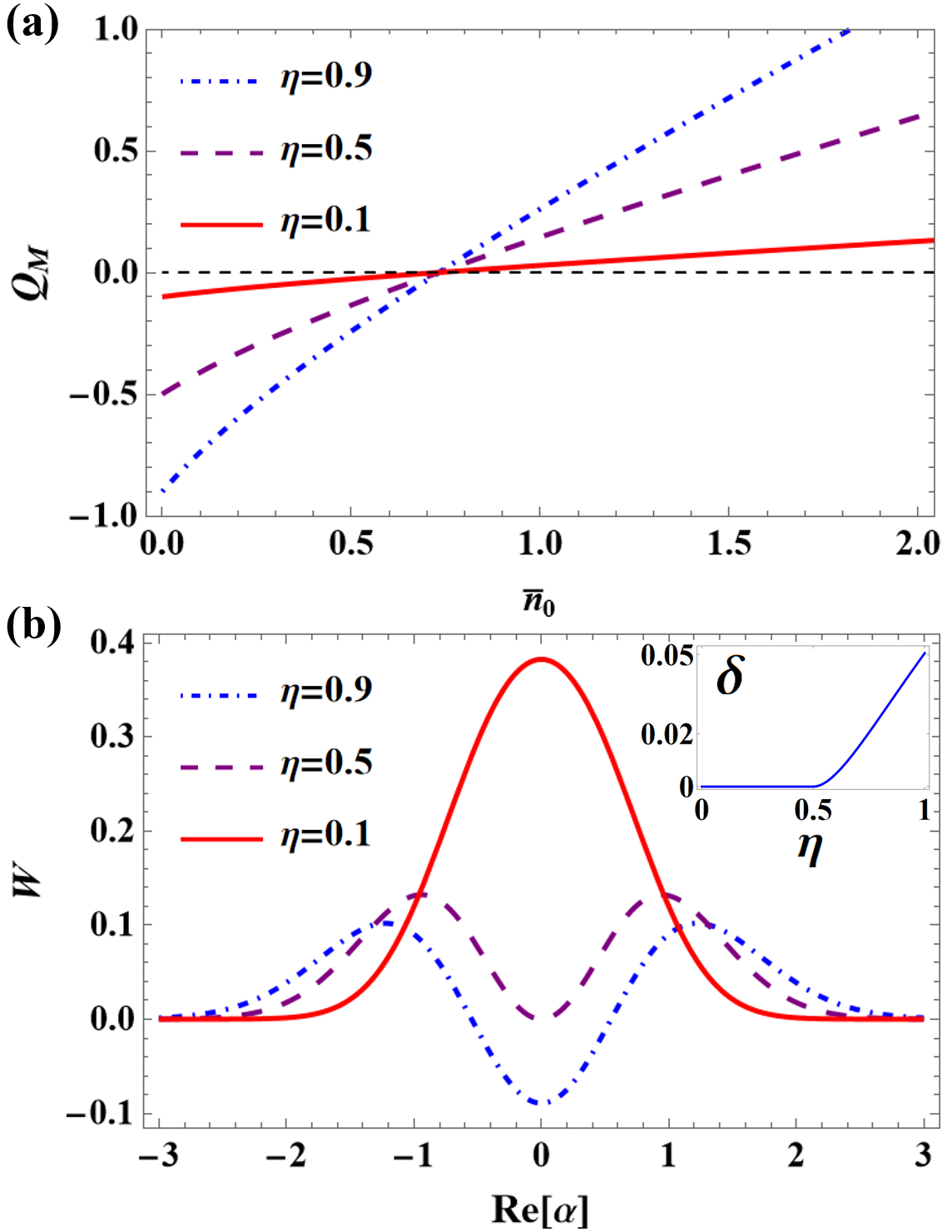}
		\caption{(a) Mandel $Q$ parameter versus the initial thermal occupation $\bar{n}_{0}$ for various state-swap efficiency $\eta$. The horizontal dashed line denotes $Q_{M}=0$ for the coherent state. (b) Wigner function [the Im$(\alpha)=0$ plane] of the cavity output field for various $\eta$. Inset shows the Wigner negativity $\delta$ versus $\eta$. We take $\bar{n}_{0}=0.8$ in (b). The other parameters are those as in Fig.~\ref{fig8}.}
		\label{fig9}
	\end{figure}

	Next, we consider the case of the single-MATS $m^\dag \rho'_{\rm th} m$. The corresponding Mandel $Q$ parameter of the microwave output field is obtained as
	\begin{equation}\label{eq:33}
		Q_{M}=\dfrac{2\eta M^{2}\bar{n}_{0}(\bar{n}_{0}+1)}{ \left[(1+M^{2})\bar{n}_{0}+1 \right] \left[(1-M^{2})\bar{n}_{0}+1 \right]}-\eta.
	\end{equation}
	In Fig.~\ref{fig9}(a), we plot $Q_{M}$ versus the initial thermal occupation $\bar{n}_{0}$ for various state-swap efficiency $\eta$. Similar to the case of the single-MACS (Fig.~\ref{fig8}(a)), a higher state-swap efficiency is preferred to see a clear sub-Poissonian character for a small $\bar{n}_{0}$. It is worth noting that the state-swap efficiency does not change the critical point,  indicated also by Eq.~\eqref{eq:33}, at which the statistics transits from sub- to super-Poissonian~\cite{Agarwal92}. 
	
	At last, we derive the Wigner function of the cavity output field of the read pulse, which reads
	\begin{equation}
		\begin{split}
			W(\alpha)&= \dfrac{2 \left(1+\bar{n}_{0}-M^{2}\bar{n}_{0} \right)^{2}}{\pi \left[1+\bar{n}_{0}+(2\eta-1)M^{2}\bar{n}_{0} \right]^{3}} e^{-2|\alpha|^{2} \frac{1+\bar{n}_{0}-M^{2}\bar{n}_{0}}{1+\bar{n}_{0}+(2\eta-1)M^{2}\bar{n}_{0}} } \\ 
			&\times \left[(\bar{n}_{0}+1) \left(4\eta|\alpha|^{2}-2\eta+1 \right) -(2\eta-1)^{2}M^{2}\bar{n}_{0} \right]. 
		\end{split}
	\end{equation}
	We plot the Wigner function for a series of $\eta$ in Fig.~\ref{fig9}(b). Similar to the single-MACS (Fig.~\ref{fig8}(c)), no negativity of the Wigner function can be found for the state-swap efficiency $\eta<0.5$, and the Wigner function behaves as a vacuum Gaussian packet when $\eta \to 0$. 

	\section{Conclusions}\label{V}
	In summary, we have shown how two kinds of magnonic NGSs, i.e., single-MACS and -MATS, can be prepared in a YIG sphere and subsequently read out by the aid of a microwave cavity. This is achieved by weakly activating the optomagnonic Stokes scattering combined with the single-photon detection. Our protocol, by adding a single magnon onto Gaussian states, provides an alternative approach for preparing a series of magnonic NGSs. These NGSs exhibit many nonclassical properties, such as sub-Poissonian statistics, quadrature squeezing, and a negative Wigner function, and may find potential applications in the fundamental studies of the quantum-to-classical transition and macroscopic quantum states, as well as in quantum information processing based on magnonics. {We note that our scheme can also be applied to prepare other non-Gaussian states, e.g., the magnonic cat-like state, which can be achieved by weakly activating the optomagnonic anti-Stokes scattering to subtract a single magnon from an initial magnonic squeezed state~\cite{Tang,Qiu20}.}

	\section*{ACKNOWLEDGMENTS}
	This work was supported by Zhejiang Provincial Natural Science Foundation of China (Grant No. LR25A050001), National Natural Science Foundation of China (Grant No. 12474365, 92265202) and National Key Research and Development Program of China (Grant No. 2024YFA1408900, 2022YFA1405200).

\end{document}